\documentclass[]{elsarticle}
\usepackage{natbib}
\usepackage{amsmath}
\usepackage{color}

\newcommand{\aver}[1]{\left\langle {#1} \right\rangle}

\begin{document}

\begin{frontmatter}

\title{Does the choice of the forcing term \\ affect flow statistics in DNS of turbulent channel flow?}

\author[polimi]{Maurizio Quadrio\corref{cor1}}
\ead{maurizio.quadrio@polimi.it}
\address[polimi]{Dipartimento di Scienze e Tecnologie Aerospaziali del Politecnico di Milano \\ 
via La Masa 34, 20156 Milano, Italy}
\cortext[cor1]{Mercator Fellow at the Karlsruhe Institute of Technology}

\author[kit]{Bettina Frohnapfel}
\ead{bettina.frohnapfel@kit.edu}
\address[kit]{Institute of Fluid Mechanics, Karlsruhe Institute of Technology (KIT) \\ 
Kaiserstr. 10, 76131 Karlsruhe, Germany}

\author[utokyo]{Yosuke Hasegawa}
\ead{ysk@iis.u-tokyo.ac.jp}
\address[utokyo]{Institute of Industrial Science, The University of Tokyo \\ 
4-6-1 Komaba, Meguro-ku, Tokyo 153-8505, Japan}

\begin{abstract}
We seek possible statistical consequences of the way a forcing term is added to the Navier--Stokes equations in the Direct Numerical Simulation (DNS) of incompressible channel flow. Simulations driven by constant flow rate, constant pressure gradient and constant power input are used to build large databases, and in particular to store the complete temporal trace of the wall-shear stress for later analysis. As these approaches correspond to different dynamical systems, it can in principle be envisaged that these differences are reflect by certain statistics of the turbulent flow field. The instantaneous realizations of the flow in the various simulations are obviously different, but, as expected, the usual one-point, one-time statistics do not show any appreciable difference. However, the PDF for the fluctuations of the streamwise component of wall friction reveals that the simulation with constant flow rate presents lower probabilities for extreme events of large positive friction. The low probability value of such events explains their negligible contribution to the commonly computed statistics; however, the very existence of a difference in the PDF demonstrates that the forcing term is not entirely uninfluential. Other statistics for wall-based quantities (the two components of friction and pressure) are examined; in particular spatio-temporal autocorrelations show small differences at large temporal separations, where unfortunately the residual statistical uncertainty is still of the same order of the observed difference. Hence we suggest that the specific choice of the forcing term does not produce important statistical consequences, unless one is interested in the strongest events of high wall friction, that are underestimated by a simulation run at constant flow rate.
\end{abstract}

\end{frontmatter}

\section{Introduction}
A flow through a straight duct always requires an external force to overcome the friction losses at the wall. In a pipeline, this force is typically generated by a pump; in a drain pipe, it is provided by gravity. Similarly, numerical simulations of duct flows also require a modelling step where this external force is defined. If we restrict our attention to the direct numerical simulation (DNS) of turbulent channel flows, where the incompressible Navier--Stokes equations are considered a more than satisfactory mathematical model, the most popular choice for the driving force is implicit in the prescription of a Constant Flow Rate (CFR): the time-dependent force is calculated by adjusting the uniform (in space) pressure gradient required to keep the flow rate constant. Alternatively, a Constant (in time) Pressure Gradient (CPG) can be prescribed, thus explicitly setting the driving force, which becomes constant and uniform. For a flow through a straight duct where friction at the wall is the only source of energy dissipation, the constant value of the pressure gradient is proportional to the mean value of wall friction, and the simulation {\em a posteriori} yields as a result the mean flow rate required to generate the prescribed frictional losses. As a third alternative besides CFR and CPG, we have recently proposed a Constant Power Input (CPI) strategy \citep{hasegawa-quadrio-frohnapfel-2014}, where the product of flow rate and pressure gradient is kept constant at every time step, while either one of them can instantaneously vary. 

For laminar flows these three simulation strategies must yield identical results, since flow rate and pressure drop are uniquely related. An impact of the forcing strategy can only be seen when the effect of changes in the flow geometry is studied, and the flow field with and without this change is to be compared. Additional frictional losses will lead to an increased pressure gradient for CFR or a decreased flow rate for CPG. The resulting difference can nicely be visualized in the money-versus-time plane \citep{frohnapfel-hasegawa-quadrio-2012}.

On the other hand, in turbulent flows the different forcing strategies lead to different time histories of the flow quantities. Such differences are generally considered to be without statistical consequences. It is generally known that the various forcing strategies are not equivalent on practical grounds; for example it is known \cite{ricco-etal-2012} that, when the value of the Reynolds number is suddenly changed in a simulation, CFR does present shorter transients and is faster to adapt to the new state of statistical equilibrium. This is a practical advantage that represents one of the reasons why the CFR approach is so far the most popular. However, the general consensus is that, once the statistical equilibrium is reached, the flow statistics do not depend upon the particular choice of the forcing term. In other words, as long as the same turbulent channel flow is considered in all cases, its statistics are independent from the forcing term.

Nevertheless, in principle a difference exists between the forcing strategies in turbulent flows. The CFR strategy results in a temporally fluctuating pressure gradient, while CPG produces a flow rate that fluctuates in time, and in CPI both quantities fluctuate. These fluctuations depend on the computational domain size and become smaller for larger domains \citep{lozanoduran-jimenez-2014}, but are always present in a domain of finite size. Owing to this difference, it can be envisaged that differences in the flow dynamics might exist, which have so far not been identified in the typically considered statistical quantities. In the present contribution, we will specifically tackle this fundamental question, by comparing various statistical quantities computed from three purposely built databases for turbulent channel flow computed with CFR, CPG and CPI at nominally the same Reynolds number. The underlying data sets are very large to ensure that errors in the statistical estimates are as small as possible. The analysis is mainly focused on the wall-shear stress, which is a quantity that can be seen as a footprint of the entire turbulent flow field above the wall. 
\section{Forcing strategies}

We consider a fully developed turbulent channel flow of an incompressible Newtonian fluid. The channel flow is set up in such a way that the streamwise coordinate corresponds to $x_1$ or $x$, the wall-normal one to  $x_2$ or $y$ and the spanwise one to $x_3$ or $z$. Throughout the entire paper dimensional quantities are marked with an asterisk, while unmarked symbols correspond to dimensionless quantities. A $+$ superscript indicates quantities made dimensionless in inner units; otherwise, the length and velocity scales are tose employed in the definition of the Reynolds number. 
 
In order to drive the fluid through the channel, the Navier--Stokes governing equations have to be supplemented by a forcing term, i.e. a force per unit volume, so that they read:
\begin{eqnarray}
\frac{\partial u_j}{\partial x_j} &=& 0 \label{eq:Cont}\\
\frac{\partial u_i}{\partial t} + u_j \frac{\partial u_i}{\partial x_j} &=& - \frac{\partial  p }{\partial x_i} + \frac{1}{Re} \frac{\partial^2 u_i}{\partial x_j \partial x_j} + F_i 
\qquad i=1,2,3
 \label{eq:NS}
\end{eqnarray}
where repeated indices imply summation, $u_i$ represents the velocity component in the $i$-th direction of the Cartesian reference  system, and $x_i$ is the corresponding spatial coordinate. The Reynolds number $Re$ is formed by the channel half-height $\delta^*$ and the characteristic velocity scale for each forcing strategy, i.e., the bulk mean velocity $U^*_b$ for CFR, the friction velocity $u^*_\tau$ for CPG or the power-based velocity $U^*_\Pi$ for CPI. As explained in \citep{hasegawa-quadrio-frohnapfel-2014}, $U^*_\Pi$ is the bulk velocity obtained in a laminar flow with the imposed power input. The additional forcing terms in the three components of the momentum equation are denoted by $F_i$. 

The present paper seeks for possible statistical consequences of selecting the streamwise component $F_1$ of the body force, the definition of which in dependence of the three forcing strategies (CPG, CFR and CPI) is described below. In respect to the other two spatial directions it is obvious that $F_2=0$, as no flow rate or pressure difference can exist between the two impermeable walls. Very often it is also considered that $F_3=0$. Strictly speaking, the two homogeneous directions are equivalent, so the question about the possible relevance of the forcing term should be asked in regard to the spanwise direction, too. However, the constant value (of the spanwise flow rate or the spanwise pressure gradient) enforced in the simulation is zero, hence possible effects, if any, are thought to be of a lesser entity. In this paper we restrict ourselves to considering $F_1$, which is labeled $F$ from here on.

\subsection{Constant Pressure Gradient (CPG)}
The simplest forcing strategy is to set $F$ to a constant value. The forcing term can be regarded as a spatially averaged pressure gradient:
\begin{equation}
\label{eq:PI}
F  = - \aver{\frac{dp}{dx}},
\end{equation}
where $\aver{ \cdot }$ indicates the spatial average in the homogeneous directions $x$ and $z$. Therefore, we refer to this strategy as CPG. Since the pressure gradient must balance the wall friction in a fully developed flow, the following force balance is satisfied:
\begin{equation}
F^* \delta^* = \tau^*_w,
\end{equation}
where $\tau^*_w$ is the mean value of the wall friction. By using the friction velocity $u^*_\tau = \sqrt{\tau^*_w / \rho^*}$, where $\rho^*$ is the density of fluid and $\delta^*$ the channel half height, the dimensionless forcing term becomes $F = 1$. The resultant Reynolds number based on $u^*_\tau$ in Eq.~(\ref{eq:NS}) is referred to as the friction Reynolds number $Re_\tau$, which is given by $Re_\tau = u^*_\tau\delta^*/\nu^*$, where $\nu^*$ is the kinematic viscosity of fluid.

\subsection{Constant Flow Rate (CFR)}
In the CFR strategy, which is often employed owing of the practical advantage mentioned above, the forcing term $F$ is not constant in time anymore, but changes at every time step in such a way that the resultant flow rate is equal to a prescribed flow rate. The flow rate per channel cross section is the bulk velocity  $U^*_b$ and the corresponding Reynolds number for CFR is thus the bulk Reynolds number $Re_b = U^*_b \delta^* / \nu^*$.

Although the CFR condition is commonly used, its detailed numerical implementation is quite often not fully described in the existing literature, and more than one variant exist which differ in the specific algorithm that keeps the flow rate constant. Sometimes the CFR strategy is detailed when the numerical method has to deal with additional complications: for example in \cite{lenormand-sagaut-taphuoc-2000} CFR is explained in the context of incompressible and compressible simulations, while in \cite{garcia-jimenez-2011} CFR is described for the simulation of an incompressible flow in a channel with riblets on the solid walls. 

Typically, the Navier--Stokes equations~(\ref{eq:NS}) is discretized as follows:
\begin{equation}
\label{eq:NS_dis}
\frac{u_i^{n+1} - u_i^{n}}{\Delta t} = {\cal N}_i^{n+\frac{1}{2}} + \delta_{i1} F^{n+1} - \left( \frac{\partial p}{\partial x_i}\right)^{n+1} + \frac{1}{2Re}\frac{\partial^2 \left( u^{n+1}_i + u^{n}_i \right) }{\partial x_j \partial x_j},
\end{equation}
where $\delta_{i1}$ is the Kronecker delta, a variable at a time step of $n$ is denoted by the superscript $n$, and ${\cal N}_i$ represents all non-linear terms in the $i$-component of the momentum equation. As the full velocity field up to time step $n$ is available, the non-linear term ${\cal N}_i^{n+\frac{1}{2}}$ is often calculated explicitly by using available information at the previous time steps, i.e. with explicit methods like Adams-Bashforth or Runge-Kutta. As for the viscous term, an implicit scheme such as the Crank-Nicolson method is often used in order to mitigate  the requirement for a small $\Delta t$ to avoid numerical instability, which becomes crucial especially in the near-wall region, where fine grid resolution is generally employed. The pressure at time step $n+1$ is often  calculated by the fractional step method. Otherwise, Eq.~(\ref{eq:NS_dis}) may be solved by a wall-normal velocity -- wall-normal vorticity formulation, where the pressure term is eliminated. In a simple flow geometry as the one considered here, when the boundary conditions are periodic, this formulation presents considerable advantages. It is important to notice, however, that the following discussion can be applied without loss of generality regardless of the numerical scheme used to solve Eq.~(\ref{eq:NS_dis}); discretization is introduced here only for establishing notation.

In the streamwise direction the forcing term $F^{n+1}$ that is required to keep the flow rate constant has to be determined. However, the  flow rate at time step $n+1$ is only known as a result of the computation. Therefore, the flow field at time step $n+1$ is assumed to be the superposition of two components, namely
\begin{equation}
\label{eq:decomposition}
u_1^{n+1} = {\hat{u}}_1^{n+1} + \tilde{u}_1^{n+1}.
\end{equation}
Here, ${\hat{u}}_1^{n+1}$ represents the solution for the streamwise velocity component at  time step  $n+1$ without application of the forcing term, i.e. is obtained with $F^{n+1}=0$. Substituting Eq.~(\ref{eq:decomposition}) into the first component of Eq.~(\ref{eq:NS_dis}) results in the following equation for the second term, $\tilde{u}_1^{n+1}$:
\begin{equation}
\label{eq:modification}
\tilde{u}_1^{n+1} = \Delta t \left( F^{n+1} + \frac{1}{2Re}\frac{d^2 \tilde{u}_1^{n+1}}{d y^2} \right).
\end{equation}
Here, we use the fact that $\tilde{u}$ is a function of $y$ only, since the forcing term does not depend on $x$ and $z$ according to its definition~(\ref{eq:PI}). The analytical solution of Eq.~(\ref{eq:modification}) can be easily obtained by taking into account the no-slip condition at a wall ($y = 0$) as:
\begin{equation}
\label{eq:modi_solution}
\tilde{u}_1^{n+1} = \Delta t F^{n+1} \left\{ 1 - \exp (-\lambda y)\right\},
\end{equation}
where $\lambda = \sqrt{2Re/ \Delta t}$. Note that this solution is essentially uniform away from the wall and only changes rapidly in the vicinity of the wall due to the viscosity. The increment of the bulk mean velocity due to $\tilde{u}_i^{n+1}$ is obtained by integrating Eq.~(\ref{eq:modi_solution}) from the wall to the channel center:
\begin{equation}
\label{eq:ub_compensate}
\Delta U_b^{n+1} = \int_0^1 \tilde{u}_1^{n+1} dy = \Delta t F^{n+1} \left( 1 - \sqrt{ \frac{\Delta t}{2Re} } \right).
\end{equation}

The numerical procedure for CFR can thus be summarized as follows. First, $\hat{u}_1^{n+1}$ is obtained by any numerical scheme with $F^{n+1}=0$. The resultant flow rate, i.e. the wall-normal integral of the $\hat{u}_1^{n+1}$ profile, now differs from the prescribed value. 
Hence, the flow rate is compensated by adding a corrective profile proportional to $\tilde{u}_1^{n+1}$ given by Eq.~(\ref{eq:modi_solution}). The amplitude of $F^{n+1}$ is determined so as to satisfy the constant flow rate condition exactly based on Eq.~(\ref{eq:ub_compensate}), which can be solved only once if the DNS integrates the equations of motion with a constant time step, whereas a condition based on the CFL number implies a variable time step and the need for solving Eq.~(\ref{eq:ub_compensate}) whenever the value of the time step is changed.

\subsection{Constant Power Input (CPI)}

Recently, the present authors have proposed \cite{hasegawa-quadrio-frohnapfel-2014} a third simulation strategy that is alternative to the CPG and CFR conditions explained above. In this strategy, it is the power input to the flow system that is kept constant in time, instead of the pressure gradient or the flow rate. The corresponding characteristic velocity $U^*_\Pi$ is one that is based on the power input $P^*_p$ to the flow system per unit wetted area: 
\begin{equation}
\label{eq:pumping_power}
P_p^* = - \aver{ \frac{dp^*}{dx^*} } \delta^* U_b^*
\end{equation}
and is given by:
\begin{equation}
U^*_\Pi = \sqrt{\frac{P^*_p \delta^*}{3 \mu^*}},
\end{equation}
where $\mu^*$ is the dynamic viscosity of fluid.
This quantity corresponds to the bulk velocity achieved by a laminar flow under a given power input $P^*_p$. In the case of turbulent flows, the resultant bulk mean velocity is smaller than $U^*_\Pi$ owing to the additional momentum loss induced by turbulent stresses. The corresponding power-based Reynolds number is defined as $Re_\Pi = U^*_\Pi \delta^* / \nu^*$.

Under the CPI condition, the product $- \aver{\frac{dp^*}{dx^*}} U^*_b$ is kept constant according to Eq.~(\ref{eq:pumping_power}). This is expressed in dimensionless form as:
\begin{equation}
P_p = - \left<\frac{dp}{dx}\right> U_b  = \frac{P^*_p}{\rho^* {U^*_b}^3} = \frac{3}{Re_{\Pi}}.
\end{equation}
so that in case of CPI the flow is driven by the fixed power based Reynolds number $Re_{\Pi}$. 

The implementation of the CPI condition to DNS is not trivial, as the power input at the next time step $P_p^{n+1}$
depends on the flow rate $U^{n+1}_b$, which is determined as a result of the computation. 
An expression for the corresponding forcing term can be obtained using the relation $F = - \aver{\frac{dp}{dx}} $ such that
$F^{n+1}$ (as introduced in Eq.~(\ref{eq:NS_dis})) can be approximated with first-order accuracy in time as:
\begin{equation}
F^{n+1} = \frac{P_p}{U^{n}_b}.
\end{equation}
The increase in the order of accuracy is straightforward, and it has been confirmed that  further increase of the accuracy does not affect the flow statistics for the small time steps typically used in DNS of turbulent channel flow.
The detailed numerical implementation of the CPI condition is described in \cite{hasegawa-quadrio-frohnapfel-2014}.
\section{Results}
Three databases have been built by carrying out three DNS of a turbulent channel flow based on the three different forcing strategies described above. The DNS code is the one described in \cite{luchini-quadrio-2006} which employs mixed spatial discretization: Fourier series in wall-parallel directions and compact finite differences in wall-normal direction. The flow is in all cases at a nominal $Re_\tau=200$, and the very same discretization (in space and time) is used. The computational domain is $4 \pi \delta^* \times 2 \delta^* \times 2 \pi \delta^*$ in the streamwise, wall-normal and spanwise directions, and matches that employed in the seminal DNS study by Kim et al \cite{kim-moin-moser-1987}. The number of Fourier modes and grid points is $N_x=N_z=256$ and $N_y=128$, with a spatial resolution of $\Delta x^+=9.6$, $\Delta y^+ = 0.8 - 4.9$ and $\Delta z^+ = 4.8$. The spatial resolution in the homogeneous directions is further increased thanks to the complete removal of aliasing error via temporary increasing the number of Fourier coefficients by at least 50\%. The time step is fixed at $\Delta t^+=0.2$, and the calculations last for 150,000 viscous time units. 750 independent flow fields for each case are stored for computing one-time statistics. When computing statistics, the two halves of the channel are averaged together to double the size of the statistical sample. The main focus of the analysis is wall friction, the two components of which are written to file for further analysis every one viscous time unit. The complete time history of wall friction thus occupies 300 GB of disk space for each case.

\begin{table}
\centering
\begin{tabular}{c c c}
forcing strategy & nominal $Re$ & measured $Re_\tau$ \\
\hline \hline
CFR & $Re_b=3173$ & $Re_\tau=199.51$ \\
CPG & $Re_\tau=200$ & $Re_\tau=199.97$ \\
CPI & $Re_\Pi=6500$ & $Re_\tau=199.71$ \\
\end{tabular}
\caption{Values of $Re$ for the three simulations. The nominal $Re$ for each simulation is shown together with the value of $Re_\tau$ measured a posteriori.}
\label{tab:cases}
\end{table}

Table \ref{tab:cases} lists, for each of the three cases, the numerical value of the specific Reynolds number that was given as input to the simulation, together with the value of $Re_\tau$ that is measured {\em a posteriori} by time averaging over the full statistical sample. It can be seen that all the cases achieve the target $Re_\tau=200$ within an extremely small tolerance. In the CPG case $Re_\tau$ is specified directly as the input value, so that the tiny difference between the target and the measured value is simply ascribed to the time-averaging process. The non-CPG cases required a little trial-and-error to identify suitable values of $Re_b$ and $Re_\Pi$, and this explains the relatively larger difference between the nominal and the actual value of $Re_\tau$. However, the present differences in $Re_\tau$ are so small (0.25\% in the worst case) that the simulations can safely be considered at the same nominal value of $Re$. The non-dimensionalization in inner units of the results presented in the following is based on the actual value of $u_\tau$ obtained for each case. 

\begin{figure}
\includegraphics[width=\textwidth]{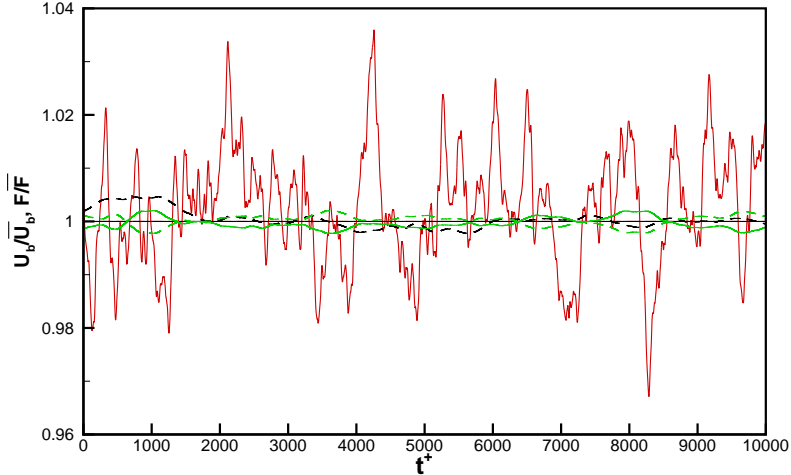}
\caption{Time histories of the bulk velocity $U_b$ (dashed lines) and the forcing term $F$ (continuous lines) with CFR (red), CPG (black) and CPI (green). Time-varying values are plotted over a relatively short time window (only 1/15 of the total length of the simulations) and are normalized by their long-term averaged values $\overline{U_b}$ and $\overline{F}$. The red dashed line is not visible as it is constant and completely overlaps the black continuous one.}
\label{fig:timehistories}
\end{figure}

Differences between CFR, CPG and CPI are visible when time traces of bulk velocity and space-mean streamwise pressure gradient (equivalent to the forcing term $F$) are compared. In general, the amplitude of the fluctuations depends upon the domain size and increases with smaller domain sizes \cite{gatti-quadrio-2013, lozanoduran-jimenez-2014}, being rather small but not negligible for the present case, where the domain is rather large as in \cite{kim-moin-moser-1987}. The forcing term itself fluctuates with CFR, the flow rate fluctuates with CPG, and both vary in time when a constant power input is prescribed in CPI. Figure \ref{fig:timehistories} shows time traces of the bulk velocity and the streamwise pressure gradient for all three cases for an arbitrarily selected and relatively short time window (only 1/15 of the total length of the simulation). The quantities are normalized with their long-term average. It can be observed that fluctuations of the pressure gradient for CFR are characterized by larger relative amplitudes and higher frequencies than fluctuations of the bulk velocity for CPG. Fluctuations of the flow rate for CPI are qualitatively similar to the ones observed for CPG. The fact that the product of flow rate and pressure gradient is constant for CPI can be easily appreciated visually by noticing that the time histories for both quantities, i.e. the two green curves, are mirror images along the horizontal axis. We also note that the temporal fluctuations of the pressure gradient and the bulk velocity in CFR and CPG, or both in CPI should all vanish if the computational box is infinitely large. In practice, however, the computational domain size is always finite, and therefore the different features of the temporal oscillation in the forcing term may affect the flow dynamics and the resultant flow statistics.


\begin{figure}
\includegraphics[width=\textwidth]{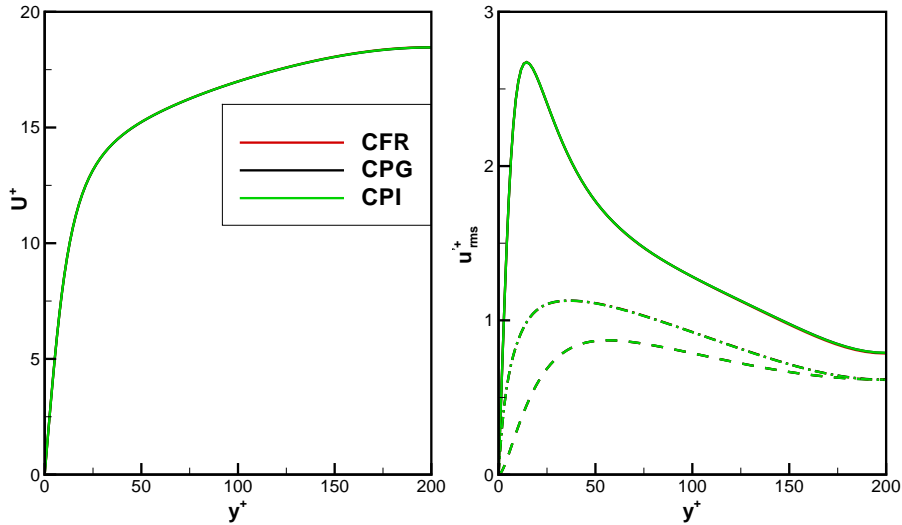}
\caption{Wall-normal profiles (in inner scaling) of the longitudinal mean velocity (left) 
and of the r.m.s. value of the fluctuations of the velocity components (right), for the three simulation strategies. In the right figure, the continuous line corresponds to the streamwise component, the dahsed line to the wall-normal one, and the dash-dotted line to the spanwise component. No difference can be appreciated.}
\label{fig:profiles}
\end{figure}

\begin{figure}
\includegraphics[width=\textwidth]{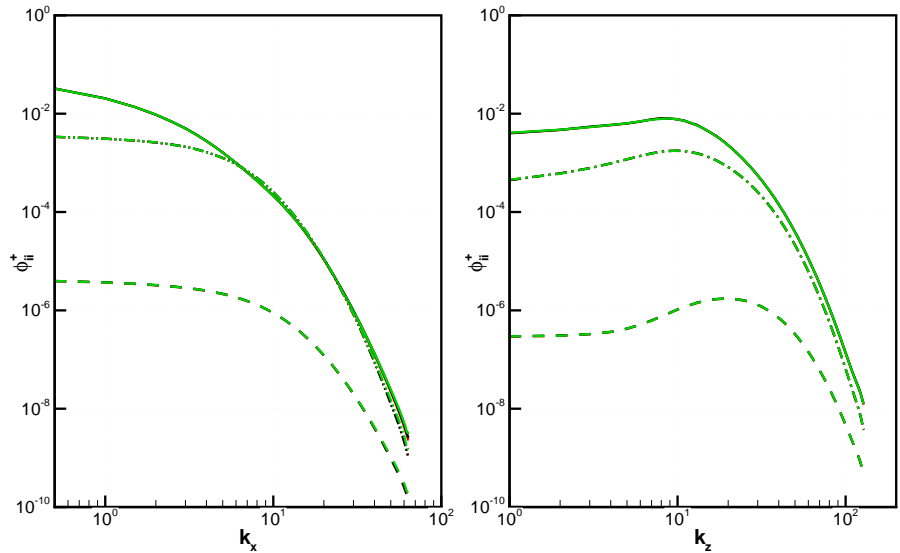}
\caption{One-dimensional spectral density functions (inner units) for the velocity components at the first inner grid point, as a function of the streamwise wavenumber $k_x$ (left) and of the spanwise wavenumber $k_z$ (right). Different lines correspond to different velocity components while the forcing strategy is coded with different colors as indicated in figure~\ref{fig:profiles}. The red and black lines cannot be seen due to the almost perfect overlap of the data.}
\label{fig:spectra}
\end{figure}

In contrast to time traces, which by themselves bear no statistical significance, the main first- and second-order single-point and single-time statistics, like the profiles of the longitudinal mean velocity or of the r.m.s. amplitude of velocity fluctuations, are expected not to present discernible differences among the three databases, in accordance to almost 30 years of practice in the DNS field. This expectation is fully confirmed. An example is shown in figure~\ref{fig:profiles}, where the three different curves for CFR, CPG and CPI are indeed undistinguishable from each other, as they overlap almost perfectly. The figure plots the mean streamwise velocity profile in linear scaling as well as the profiles of the r.m.s. values of velocity fluctuations. Similar results hold for other such statistics, among which we only show in figure \ref{fig:spectra} one-dimensional spectral density functions for the velocity components in the first inner grid point, with the sole aim of showing their smoothness, made possible by the sheer size of the databases.

\begin{table}
\centering
\begin{tabular}{c c c c}
forcing strategy & r.m.s. & third mom. & fourth mom. \\
\hline \hline
CFR & 0.3719 & 0.9405 & 4.2924 \\
CPG & 0.3720 & 0.9380 & 4.2895 \\
CPI & 0.3720 & 0.9400 & 4.2945 \\
\end{tabular}
\caption{Statistical moments for the streamwise component of wall friction. The second moment is expressed as the root-mean-squared value of the fluctuations divided by the mean value. Third and fourth moments are by definition dimensionless. At $Re_\tau=180$ values of 0.367, 0.930 and 4.219 are calculated in \cite{lenaers-etal-2012}.}
\label{tab:tau-stats}
\end{table}

\begin{figure}
\includegraphics[width=0.49\textwidth]{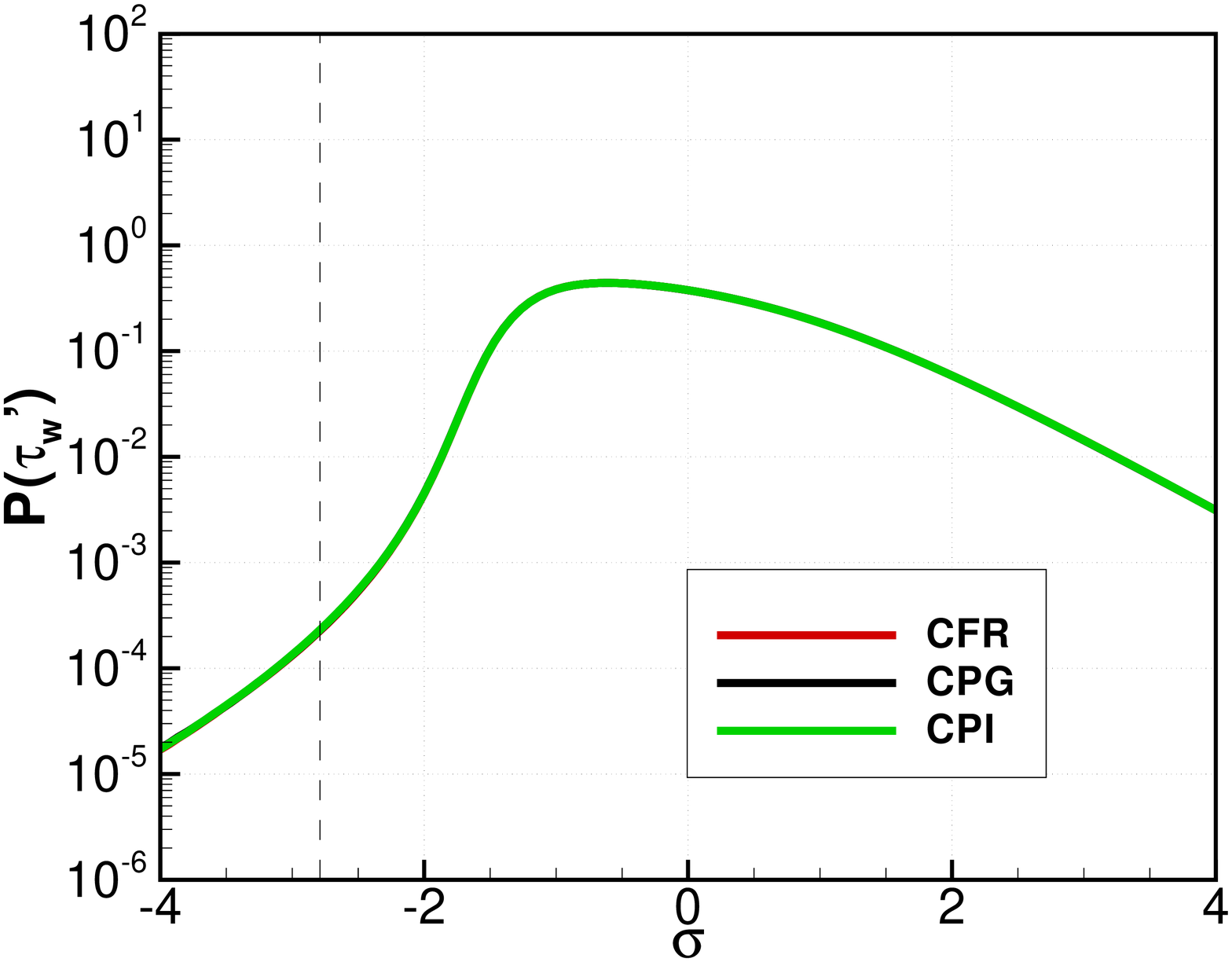}
\includegraphics[width=0.49\textwidth]{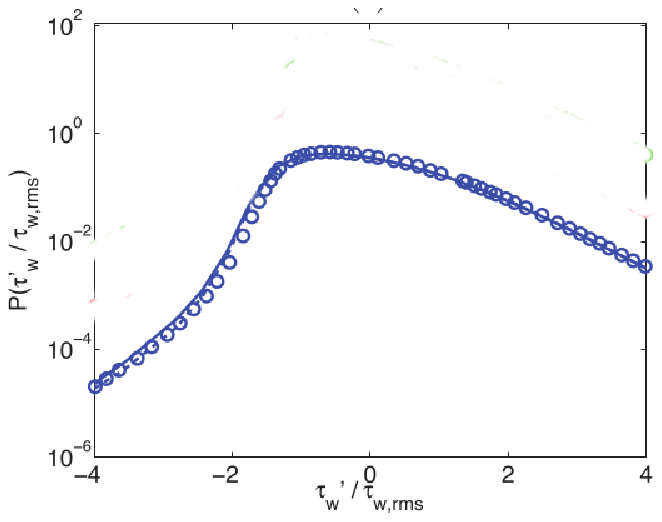}
\caption{Probability density function of the instantaneous, local fluctuations of the streamwise component of wall friction around their time- and space-mean value. Abscissa is in units of standard deviation. Shown are the present results (left) and results taken from \cite{lenaers-etal-2012} at the slightly lower $Re_\tau=180$ (right). For the present data negative local friction corresponds to $\sigma < -2.79$, marked by the vertical dashed line. No difference between the three forcing strategies can be appreciated.}
\label{fig:tau-comparison}
\end{figure}

For further analysis, we  focus in particular on the statistics of streamwise wall friction, for which we have stored full time histories and not only uncorrelated flow fields. As indicated by the Reynolds numbers reported in Table \ref{tab:cases}, all three data sets have an almost equivalent friction Reynolds number of $Re_\tau=200$ which indicates a nearly identical mean value of wall friction. This is confirmed by the values shown in Table \ref{tab:tau-stats}, where additionally the r.m.s. value of the fluctuations expressed in percentage of the mean is given, together with the dimensionless skewness and flatness factors. All of them agree very well among themselves as well as with available published data \cite{lenaers-etal-2012}.

It has recently been recognized that rare events in the near-wall flow dynamics may lead to localized backflow, hence to negative values of streamwise friction. Lenaers et al \cite{lenaers-etal-2012} made this observation in turbulent channel flow, but similar results are also observed in the zero pressure gradient turbulent boundary layer \cite{schlatter-orlu-2010, schlatter-etal-2010}. By looking at the probability density function for the wall friction, it was determined in \cite{lenaers-etal-2012} that the probability of local backflow is around 0.01\% at $Re_\tau=180$, and has a tendency to increase with $Re$. Here we compute the same PDF and confirm that result; actually a slightly higher probability of about 0.02\% for backflow is observed, which sounds reasonable given the slightly larger value of $Re$ considered here ($Re_\tau=200$ vs $Re_\tau=180$). Overall, the PDF for the three forcing types look very similar to the one reported by \cite{lenaers-etal-2012}, without revealing any difference related to the forcing strategy.

\begin{figure}
\includegraphics[width=\textwidth]{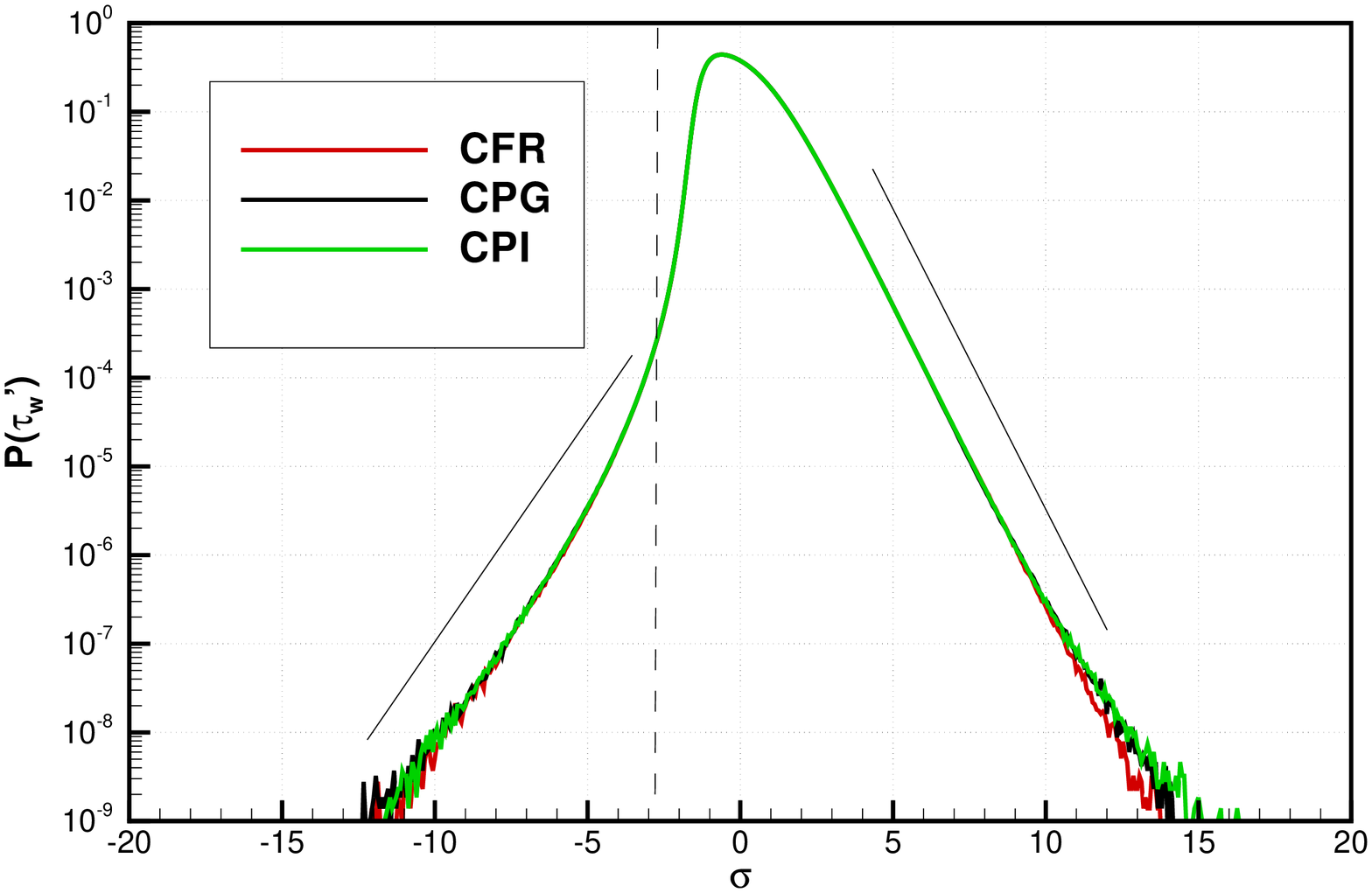}
\caption{Probability density function  of the instantaneous, local fluctuations of the streamwise component of wall friction around their time- and space-mean value. The abscissa is in units of standard deviation. Negative local friction is found for $\sigma < -2.79$, marked by the vertical dashed line. The two oblique straight lines emphasize the exponential character of the PDF tails.}
\label{fig:full-PDF}
\end{figure}

A different scenario emerges when the same PDF is plotted by taking advantage of the large size of the database, that comprises several billions datapoints. Such a database allows extreme events to be tracked, and the PDF plot can be extended to contemplate much lower probability values. This full plot is shown in figure \ref{fig:full-PDF}, where the PDF is observed to extend for almost nine decades. Much wider tails are evident now, with the tails being remarkably exponential (straight lines in this plot), especially for $\sigma>0$. The PDF are obviously quite irregular in the extreme part of their tails, owing to the small number of events belonging there. Nevertheless, the forcing term, while apparently not affecting the extreme events of negative friction, does indeed appear to affect the strongest wall shear stress events with positive friction. In particular, CFR presents less of such very rare events. The probability is extremely low in all cases, as we are speaking of events with a probability of occurrence of $10^{-8}-10^{-9}$. However, the probability of observing strong events with $\sigma>12$ is 2--3 times lower for CFR than for the other two forcing strategies.

\begin{figure}
\includegraphics[width=\textwidth]{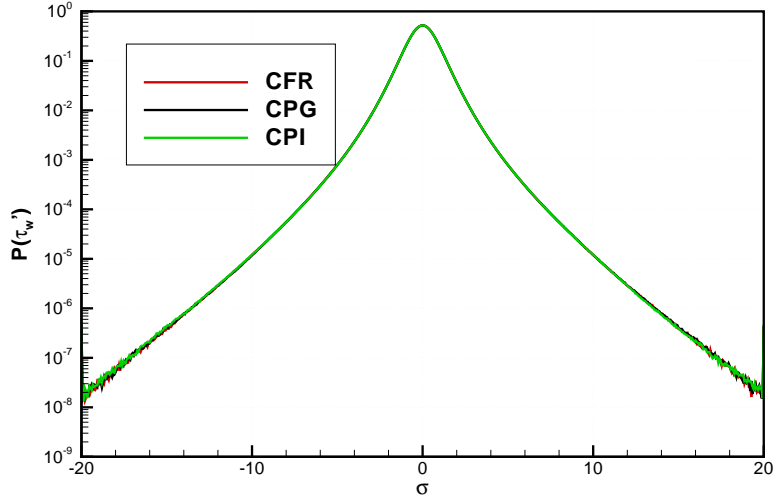}
\caption{PDF of the instantaneous, local fluctuations of the spanwise component of wall friction around the time- and space-mean value. The abscissa is in units of standard deviation.}
\label{fig:full-PDF-tauz}
\end{figure}

As shown in Table~\ref{tab:cases}, the actual friction Reynolds numbers of the three cases are slightly different. In particular, the Reynolds number of CFR is the smallest among the three cases, although by a tiny amount. Therefore, one might suspect that the present deviation is caused by the small difference in $Re_\tau$. Indeed, it is known \cite{orlu-schlatter-2011} that there is a slow trend of increase for such quantities with $Re$. However, if this were the case a similar difference would also be expected to show up for the the spanwise wall shear stress.  Since the PDFs of the spanwise wall shear stress as shown in figure \ref{fig:full-PDF-tauz} do not show discernible differences among the three cases, a Reynolds number dependency within the results presented in figure \ref{fig:full-PDF} may be ruled out. The PDF for the spanwise component of the wall friction (figure \ref{fig:full-PDF-tauz}) are symmetrical due to the reflectional symmetry of the problem ($z \rightarrow -z$), and the three curves for the different forcing strategies basically overlap. Considering that the forcing term in CFR shows the largest fluctuations as shown in figure \ref{fig:timehistories}, it might be expected that this directly increases the fluctuations of the spatially-averaged wall shear stress. However, figure \ref{fig:full-PDF} indicates that the extreme events of the very lare wall shear stress are less probable in CFR. Hence, the difference cannot be explained as a direct consequence of the forcing and therefore suggests the existence of an indirect dynamical effect on near-wall turbulence.


\begin{figure}
\includegraphics[width=\textwidth]{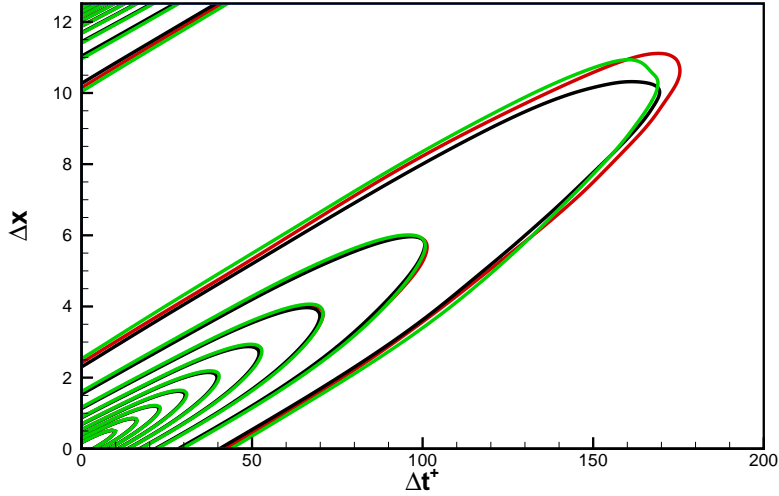}
\caption{Space-time autocorrelation function $R$ of the wall friction at zero spanwise separation. Lines show levels of the autocorrelation functions from 0.9 down to 0.1 by 0.1 decrements. The color code is the same as used before; i.e. red=CFR, black=CPG, green=CPI.}
\label{fig:comparison}
\end{figure}

We now move to two-point, two-times statistics, and show in figure \ref{fig:comparison} 
the autocorrelation $R$ of the wall friction as a function of the spatial (longitudinal) and temporal correlation, computed at zero spanwise separation. The definition of $R$ is
\begin{equation}
R(\Delta x, \Delta t) = \frac{ \aver{ \tau'_w(x, t) \tau'_w(x+\Delta x, t + \Delta t)} }{ {\tau'_w}^2_{rms} },
\end{equation}
where a prime denotes a fluctuating component from the time-space average.
The autocorrelation functions show the well-known "banana" or "cigar" shape, that indicates the convective nature of the flow: the slope of the inclined narrow region, which dimension-wise represents a velocity in such a plane, broadly corresponds to the convection velocity of the friction fluctuations \cite{kim-hussain-1993, quadrio-luchini-2003} whose value is around 10 when expressed in inner units. The autocorrelation of the wall friciton has been looked at in the past in several papers, but mainly as a function of time separation alone, or space separation alone, i.e. along one of the axes of figure \ref{fig:comparison}. Indeed, temporal correlations are the obvious choice for experimentalists; spatial correlations are easily computed from DNS, whereas exploring the correlation in the space-time plane requires the database to be fully time resolved. The three correlation functions computed here are almost identical in the whole small-separation region where $R$ assumes large values, say $R>0.2$. However, the contours at $R=0.1$ do indeed show some difference that could be ascribed to something more than residual error. 

\begin{figure}
\includegraphics[width=\textwidth]{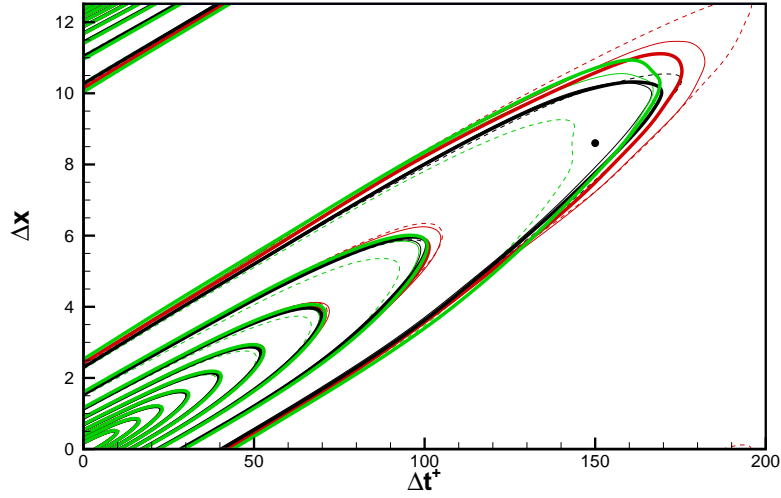}
\caption{Space-time autocorrelation function $R$ of the wall friction at zero spanwise separation. Thick continuous lines are computed by using the full statistical sample, dashed lines are computed by using the first one-third of the dataset, and thin continuous lines are computed using the first two-thirds of the dataset. The dot indicates where the convergence process of the autocorrelation function is monitored (see next figure \ref{fig:convergence}). The color code is the same as used before; i.e. red=CFR, black=CPG, green=CPI.}
\label{fig:uncertainty}
\end{figure}

\begin{figure}
\includegraphics[width=\textwidth]{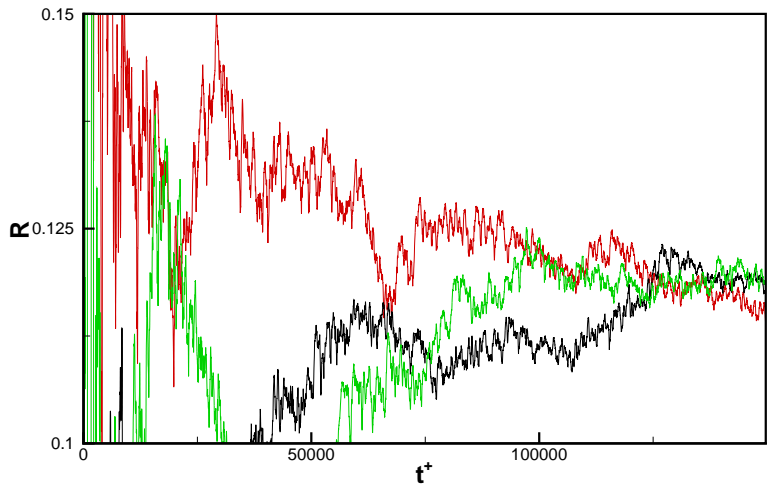}
\caption{Convergence history of the correlation function $R$ for $\Delta x=8.6$ and $\Delta t^+=150$. The color code is the same as used before; i.e. red=CFR, black=CPG, green=CPI.}
\label{fig:convergence}
\end{figure}

This difference was much larger when only a portion of the database was available. To investigate this, a simple visual estimate of how the curves in the high-separation region are converging upon increasing time average can be obtained in figure \ref{fig:uncertainty}, where the same correlation function of figure \ref{fig:comparison} is replotted by only using a partial chunk of the whole database. It becomes evident that the contours at $R=0.1$ present some residual variations upon increasing the statistical sample, and the possibility exists that they eventually converge upon further increasing the size of the sample. This view is better supported by figure \ref{fig:convergence}, where the convergence process of the value of the autocorrelation for a fixed separation of $\Delta t^+ = 150$ and $\Delta x=8.6$ (shown with a black dot in figure \ref{fig:uncertainty}) is monitored as the computing time progresses. The very slow convergence process at such large temporal separations can be appreciated, together with the tendency of the three curves to progressively approach the same value. 

\begin{figure}
\includegraphics[width=\textwidth]{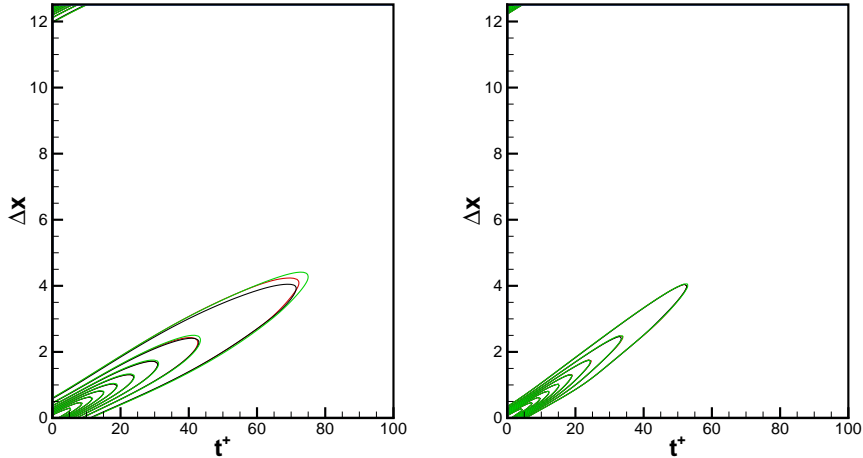}
\caption{Two-dimensional autocorrelation of the spanwise component of wall friction (left) and pressure (right) at zero spanwise separation. The color code is the same as used before; i.e. red=CFR, black=CPG, green=CPI.}
\label{fig:others}
\end{figure}

To conclude, we show in figure \ref{fig:others} the same spatio-temporal correlations for the other quantities that are defined at a solid wall, i.e. the spanwise component of the wall shear (shown in the left plot) and the pressure (right plot). The contours for both quantities are qualitatively similar to those for the streamwise component of wall shear, with the elongation implied by the convective nature of the flow and the convective velocity scale shown by the mean slope of the contours. The typical Lagrangian timescale associated to these correlations is shorter than that for the longitudinal friction, and the convective velocity scale is larger, especially for the pressure correlation. What is most relevant here, however, is that neither function shows differences that can be attributed to the particular forcing strategy. Indeed, the spanwise friction shows some differences at the lowest correlation level $R=0.1$, that may be due to a still incomplete convergence as in the case of longitudinal friction.

\section{Concluding remarks}
Since the choice of a specific forcing term in the DNS of a turbulent channel flow is to some extent arbitrary, in this paper we have searched for possible statistical consequences of this choice. It is not inconceivable that such signature exists, as different forcing terms produce different dynamical systems, unless an infinitely large computational domain is considered.

The difference between the forcing terms for CFR, CPG and CPI can indeed be seen in the time traces of instantaneous global flow properties. In order to obtain an indication whether these differences are also reflected in some statistical properties of the flow, three databases have been built by DNS, one for each forcing strategy. 

As far as one-point one-time statistics are concerned, the agreement among the results of the three forcing strategies is extremely good. This was the expected outcome, as extensive DNS research over several decades did never identify any effect of the forcing term in such frequently encountered statistics. However, we have observed small yet statistically significant differences in the PDF of the streamwise wall friction, where CFR shows fewer very rare events of extremely high wall friction. For such differences to become evident, a very large database is required as the PDF begin to deviate for very rare events, whose intensity is $\sigma > 12$. In other words, in such events the amplitude of the friction fluctuations above the mean is almost 5 times the mean value, and their probability of occurrence is about $10^{-8}$. Small differences might also be hidden in the space-time autocorrelation of wall friction, although this quantity is more sensitive to statistical error and the size of our database is still not large enough to enable a firm conclusion on this matter. 

Hence, we have found no reason for questioning the way DNS research of turbulent flows is carried out, i.e. by choosing whatever forcing term turns out to be the most convenient in practice. However, having found at least one statistical quantity (the wall friction PDF) that is affected by the specific forcing term is an important result in principle. It can also be envisaged that other quantities too might carry traces of the forcing term, or that differences might be observed in a Lagrangian framework. In general, these differences are likely to be more pronounced as the size of the computational domain is reduced. For sure, they matter whenever a flow is changed from one state to another, which is typical, for example in flow control. The choice of the forcing term might also be important in studies on laminar to turbulent transition.  

\section*{Acknowledgments}
Support through DFG project FR 2823/5-1 is gratefully acknowledged. 
Y.H. acknowledges the support by the Ministry of Education, Culture, Sports, Science and Technology of Japan (MEXT) through 
the Grant-in-Aid for Scientific Research (B) (No. 25289037)

\bibliographystyle{elsarticle-harv}

\end{document}